\documentclass[twocolumn,tighten]{aastex62}
\usepackage{xspace}
\newcommand{\msun}{\ensuremath{\textrm{M}_\odot}\xspace}
\newcommand{\ammo}{\ensuremath{\textrm{NH}_3}\xspace}

\newcommand{\kms}{{\ensuremath{{\rm km\, s^{-1}}}}\xspace}
\newcommand{\kmspc}{{\ensuremath{{\rm km\,s^{-1}\,pc}}}\xspace}

\shorttitle{Specific angular momentum profile}
\shortauthors{Pineda et al.}
\begin{document}

\title{The specific angular momentum radial profile in dense cores: \\
improved initial conditions for disk formation}

\author[0000-0002-3972-1978]{Jaime E. Pineda}
\affiliation{Max-Planck-Institut f\"ur extraterrestrische Physik, Giessenbachstrasse 1, 85748 Garching, Germany}
\email{jpineda@mpe.mpg.de}

\author[0000-0002-5359-8072]{Bo Zhao}
\affiliation{Max-Planck-Institut f\"ur extraterrestrische Physik, Giessenbachstrasse 1, 85748 Garching, Germany}
\author[0000-0002-1730-8832]{Anika Schmiedeke}
\affiliation{Max-Planck-Institut f\"ur extraterrestrische Physik, Giessenbachstrasse 1, 85748 Garching, Germany}

\author[0000-0003-3172-6763]{Dominique M. Segura-Cox}
\affiliation{Max-Planck-Institut f\"ur extraterrestrische Physik, Giessenbachstrasse 1, 85748 Garching, Germany}

\author[0000-0003-1481-7911]{Paola Caselli}
\affiliation{Max-Planck-Institut f\"ur extraterrestrische Physik, Giessenbachstrasse 1, 85748 Garching, Germany}

\author[0000-0002-2885-1806]{Philip C. Myers}
\affiliation{Harvard-Smithsonian Center for Astrophysics, 60 Garden St., Cambridge, MA 02138, USA}

\author[0000-0002-6195-0152]{John J. Tobin}
\affiliation{National Radio Astronomy Observatory, 520 Edgemont Rd., Charlottesville, VA 22901, USA}

\author[0000-0003-0749-9505]{Michael Dunham}
\affiliation{Department of Physics, State University of New York at Fredonia, 280 Central Avenue, Fredonia, NY 14063, USA}
\affiliation{Harvard-Smithsonian Center for Astrophysics, 60 Garden St., Cambridge, MA 02138, USA}

\begin{abstract}
The determination of the specific angular momentum radial profile, $j(r)$, in the early stages of star formation is crucial to constrain star and circumstellar disk formation theories. The specific angular momentum is directly related to the largest Keplerian disk possible, and it could constrain the angular momentum removal mechanism. We determine $j(r)$ towards two Class 0 objects and a first hydrostatic core candidate in the Perseus cloud, which is consistent across all three sources and well fit with a single power-law relation between 800 and 10,000\,au:  $j_{fit}(r)=10^{-3.60\pm 0.15}\left(r/\textrm{1,000\,au}\right)^{1.80\pm0.04}\,\kmspc$. This power-law relation is in between solid body rotation ($\propto r^2$) and pure turbulence ($\propto r^{1.5}$). This strongly suggests that even at 1,000\,au, the influence of the dense core's initial level of turbulence or the connection between core and the molecular cloud is still present. The specific angular momentum at 10,000\,au is $\approx3\times$ higher than previously estimated,  while at 1,000\,au it is lower by $2\times$. We do not find a region of conserved specific angular momentum, although it could still be present at a smaller radius.
We estimate an upper limit to the largest Keplerian disk radius of 60\,au, which is small but consistent with published upper limits.
Finally, these results suggest that more realistic initial conditions for numerical simulations of disk formation are needed.  Some possible solutions include: a) use a larger simulation box to include some level of driven turbulence or connection to the parental cloud, or b) incorporate the observed $j(r)$  to setup the dense core kinematics initial conditions.
\end{abstract}


\section{Introduction}

What is the role of angular momentum in disk formation?
Currently there are no good observational constraints on how and when a circumstellar disk, which later will form planets, gathers most of its mass. 
Three small surveys have attempted to study the presence and properties of these disks in the Class 0 stage, the earliest stages of protostellar evolution. 
Two of them \citep{J_rgensen_2009,Enoch_2011}, found evidence for massive disks (0.02--2\msun) around low-mass stars and concluded that disks universally form and gather material early in the Class 0 stage. With linear resolutions ranging from 200--1,000 au, neither survey was capable of resolving disks and instead inferred their presence from compact, unresolved components in the visibilities (in Fourier space instead of image space). 
Furthermore, both surveys were biased toward fairly luminous sources and are not representative of the full range of protostellar luminosities. 
Recent observations have shown clear evidence of disks towards a few Class 0 objects \citep{Murillo_2013,Ohashi_2014}. 
However,  Plateau de Bure Interferometer (PdBI) observations of 5 Class 0 sources found no evidence for disks in most of them \citep{Maury_2010}, 
therefore suggesting that disks gather most of their mass at later stages. 
\cite{Dunham_2014} showed that at least 50\% of the protostellar disks 
observed to date are consistent with the range of disk masses found in  
non-magnetic hydrodynamical simulations. 
The uncertainty concerning disk properties during the Class 0 stage is not limited to observational studies. 
A similar disagreement over whether large and massive disks are possible during the Class 0 stage is found in numerical simulations. 
The debate boils down to the mechanisms of transporting angular momentum  and their efficiency. 
Magnetic braking can efficiently transport angular momentum away from the central/disk-forming region, and therefore strongly suppress disk formation, generating small (few au) and low mass disks during the Class 0 stage \citep[e.g.][]{Allen_2003,Hennebelle2008,Mellon_2008,Seifried_2011}. 
However, many other potential solutions have been proposed to reduce the effect of magnetic breaking and allow disks to form, including magnetic flux loss through various mechanisms (non-ideal MHD effects), outflow-induced envelope clearing, misaligned magnetic field, turbulence, and dust evolution  \citep{Mellon_2008,Joos_2012,Seifried_2012,Li_2014,Tomida_2013,Tomida_2015,Zhao2016}. 
These scenarios allow the formation of sizeable (tens of au) and massive disks during the Class 0 stage. 
Massive disks are expected to promptly fragment to form binary and multiple systems in this early stage of evolution \citep{Zhao2018}.
Future high-angular resolution observations will determine how common (or not) big disks are during the Class 0 stage \citep[e.g.][]{Segura-Cox_2018,Maury_2019-Calypso}. 
However, this will not address the initial conditions or physics needed to reproduce observations. 
In particular, the specific angular momentum of the infalling material is directly related to the possible protostellar disk radius  and how much angular momentum must be transported away.

A determination of the specific angular momentum as a function of radius, starting from the scale of dense cores down to their inner envelope (10,000\,au down to 1,000\,au), has been challenging. 
Estimation of the specific angular momentum for dense cores have mostly been done on the largest scales, $\sim$0.1\,pc \citep{Goodman_1993,Caselli_2002}, while estimates at smaller scales have been done with objects with disks \citep[Class 0/I,][]{Ohashi1997,Chen_2007,Tobin_2012,Kurono2013,Yen_2015} and on the prestellar core L1544 \citep{Crapsi_2007-L1544}. 
From these heterogeneous ensemble of measurements it is suggested \citep{Ohashi1997,Belloche_2013} that some of the specific angular momentum is lost from the largest scales, while at scales smaller than 
$\sim$5,000 au the specific angular momentum is conserved until reaching the scales of disks ($\sim$100 au). 
A recent survey towards 17 Class 0/I objects using the SMA \citep{Yen_2015} determined that the specific angular momentum, $j$, is between $10^{-5}$--$4\times10^{-3}$~\kmspc at 1,000\,au. 
Recently, \cite{Yen_2015b} estimated the specific angular momentum of B335 to be 
$4.3\pm 0.5\times 10^{-5}$~\kmspc (1.3$\times$10$^{19}$\,cm$^2$\,s$^{-1}$) at $\sim$180\,au, using high-resolution ALMA observations.
These results show that the specific angular momentum present a large scatter, with values much lower than those previously suggested \citep{Ohashi1997,Belloche_2013}. 
In addition, numerical simulations (with and without magnetic fields) need to include some initial angular momentum to form a disk. 
While the total angular momentum is somewhat constrained, its radial distribution is not. 
Therefore, the determination of the angular momentum radial profile and its dependence on the environment will be crucial in understanding disk formation.
Recently, \cite{Tatematsu_2016} carried out a similar analysis on the cores in Orion~A using single dish low-resolution N$_2$H$^+$ (1--0) maps, and  they find similar trends to \cite{Goodman_1993}, although with marginally larger $J/M$, were $J$ and $M$ are the total angular momentum and mass of the core.

Another path to address this question is to determine the specific angular momentum radial profile from the dense core scales ($\sim$10,000\,au) 
down to the scales relevant to disk properties ($\sim$50\,au). 
In this paper, we present the results of determining the specific angular momentum radial profile from 
$\sim$10,000\,au down to $\sim$1,000\,au towards three of the youngest sources in the Perseus molecular cloud.
These sources are close to edge-on or with an elongated morphology in the plane of the sky (and are likely tracing the inner flattened envelope):
two Class~0 objects --- HH211 and IRAS03282+3035 --- 
and one First Hydrostatic Core candidate, L1451-mm.
Finally, we discuss the relevance of the results for numerical simulations of disk formation. 

\section{Data}

We use archival Very Large Array (VLA) interferometric observations of the \ammo(1,1) line at 23.694\,GHz 
in the compact (D) configuration. 
All sources are in the Perseus molecular cloud \citep[300\,pc,][]{Zucker_2018-Perseus_distance,Ortiz-Leon2018_Perseus-Distance} 
and the achieved spectral resolution  is 0.154\,\kms. 
This line is a good tracer of the dense gas in cores \citep{Benson_1989,Goodman_1998,Tafalla_2004,Pineda_2010,Pineda2015-Multiple}.
The data presented here have already been published \citep{Tanner_2010,Pineda_2010,Tobin_2011}, although here we usually obtain smaller beam sizes than those 
used in the original papers since we are not trying to detect the \ammo(2,2) line to derive kinetic temperatures, which allow us to probe smaller scales at the 
price of lower sensitivity for faint extended emission. 
Details of the observations are listed in Table~\ref{table:obs}.

\subsection{HH211}
The observations were carried out on 2005 December 13 under project AA300. 
The original analysis of the data is presented by \cite{Tanner_2010}.
The data reduction and imaging were carried out using CASA 4.4.0  \citep{2007ASPC..376..127M}. 
Imaging was done using a robust parameter of 0.5 and multiscale clean, which significantly reduces 
the presence of artifacts. 
Multiscale clean is used with scales of 0, 3, 9, and 27~arcsec. 
The beam size and rms levels (estimated from line free channels) are reported in Table~\ref{table:obs}.

\subsection{L1451-mm}
The observations were carried out on 2006 January 11 under project AA300. 
The original analysis of the data is presented by \cite{Pineda_2011}.
The data reduction and imaging were carried out using CASA 4.4.0 \citep{2007ASPC..376..127M}. 
Imaging was done using a robust parameter of 0.5 and multiscale clean, which significantly reduces 
the presence of artifacts. 
Multiscale clean is used with scales of 0, 3, 9, and 27 arcsec. 
The beam size and rms levels (estimated from line free channels) are reported in Table~\ref{table:obs}.

\subsection{IRAS03282+3035 (IRAS03282)}
The observations were carried out on 2009 November 11 under project AT373. 
The original analysis of the data is presented by \cite{Tobin_2011}.
The data reduction and imaging were carried out using CASA 4.4.0 \citep{2007ASPC..376..127M}. 
Imaging was done using a robust parameter of 0.5 and multiscale clean, which significantly reduces 
the presence of artifacts. 
Multiscale clean is used with scales of 0, 3, 9, and 27 arcsec. 
The beam size and rms levels (estimated from line free channels) are reported in Table~\ref{table:obs}.

\floattable
\begin{deluxetable*}{ l c c c c c c}
\tablecaption{Parameters of Interferometric Maps\label{table:obs}}
\tablewidth{0pt}
\tablehead{\colhead{Source} &\colhead{Project code} & \colhead{Observing date} & \colhead{RA} & \colhead{Dec} & \colhead{beamsize} & \colhead{rms} \\ 
\colhead{} & \colhead{} & \colhead{(YYYYMMDD)} & \colhead{(hh:mm:ss.ss)} & \colhead{(dd:mm:ss.s)} & \colhead{(PA)} & \colhead{(mJy beam$^{-1}$)}}
\startdata
        IRAS03282 & AT373 & 20091111 & 3:31:20.94 & 30:45:30.3 & $3.\arcsec63\times$2.\arcsec75 (89.56  deg) & 2.5\\
        HH211        & AA300 & 20051213 & 3:43:56.52 & 32:00:52.8 & 2.\arcsec63$\times$2.\arcsec55 (-55.54 deg) & 2\\
        L1451-mm  & AA300 & 20060111 &3:25:10.21 & 30:23:55.3 & 3.\arcsec09$\times$2.\arcsec61 (-19.79 deg) & 3\\
\enddata
\end{deluxetable*}

\begin{deluxetable*}{lccccccc c}
\tablecaption{Protostar  parameters\label{table:yso}}
\tablewidth{0pt}
\tablehead{\colhead{Source} & \colhead{RA} & \colhead{Dec} 
& \colhead{Disk Inclination\tablenotemark{a}}  & \colhead{Outflow PA\tablenotemark{b}} &
\colhead{Disk radius} & \colhead{$T_{bol}$} & \colhead{Other names} & \colhead{Binary}\\ 
\colhead{} & \colhead{(hh:mm:ss.ss)} & \colhead{(dd:mm:ss.ss)} & 
\colhead{(deg)} & \colhead{(deg)} & \colhead{(au)} & \colhead{(K)} & \colhead{} & \colhead{}}
\startdata
        IRAS03282 & 3:31:20.94 & 30:45:30.27 & Unconstrained & 122 & $<20$\tablenotemark{c} & 33 & Per-emb-5\ & Maybe\tablenotemark{d} \\ 
        HH211        & 3:43:56.81 & 32:00:50.20 & 51 & 117 & 12\tablenotemark{c} & 24 & Per-emb-1 & No \\ 
        L1451-mm  & 3:25:10.21 & 30:23:55.30 & $>64$ & 10 & $<107$ & $<$30&  & No \\ 
\enddata
\tablenotetext{a}{Measured from the plane of the sky, $i=0\deg$ is edge-on.}
\tablenotetext{b}{Measured East from North.}
\tablenotetext{c}{Disk radius estimate from \cite{Segura-Cox_2018}, rescaled to the current best estimate for Perseus of 300\,pc, which are consistent with the estimates from \cite{Lee_2018-HH211} and \cite{Lee_2019-HH211}.}
\tablenotetext{d}{A 20\,au separation companion is identified in the VLA 9\,mm continuum map \citep{Tobin_2016-VANDAM_Multiple}, however, a re-assessment on the companionship and disk properties based on ALMA data might change some of these results (Tobin et al., in prep.)}
\end{deluxetable*}

\section{Results}

\subsection{Line fit}
We fit the NH$_3$ (1,1) line with all the hyperfine components using the \verb+pyspeckit+ package \citep{2011ascl.soft09001G}. 
This implements the \ammo hyperfine structure fit described in \cite{Rosolowsky_2008} within \verb+python+.
This method minimizes the difference between the observed profile and a synthetic \ammo line profile 
that is parametrized by: the line centroid velocity ($V_{lsr}$), velocity dispersion ($\sigma_v$), 
kinetic temperature ($T_{kin}$), excitation temperature ($T_{ex}$), and column density ($N_{{\rm NH_3}}$).

Here we fix the kinetic temperature to 12\,K, since we do not fit the weaker \ammo(2,2) lines, 
and therefore the derived column densities are unconstrained. 
However, this does not affect the analysis since it only relies on the kinematic properties of the gas, 
which are well constrained even when \ammo(2,2) is undetected \citep{Friesen2017}.

\begin{figure*}
\begin{center}
\includegraphics[width=\textwidth]{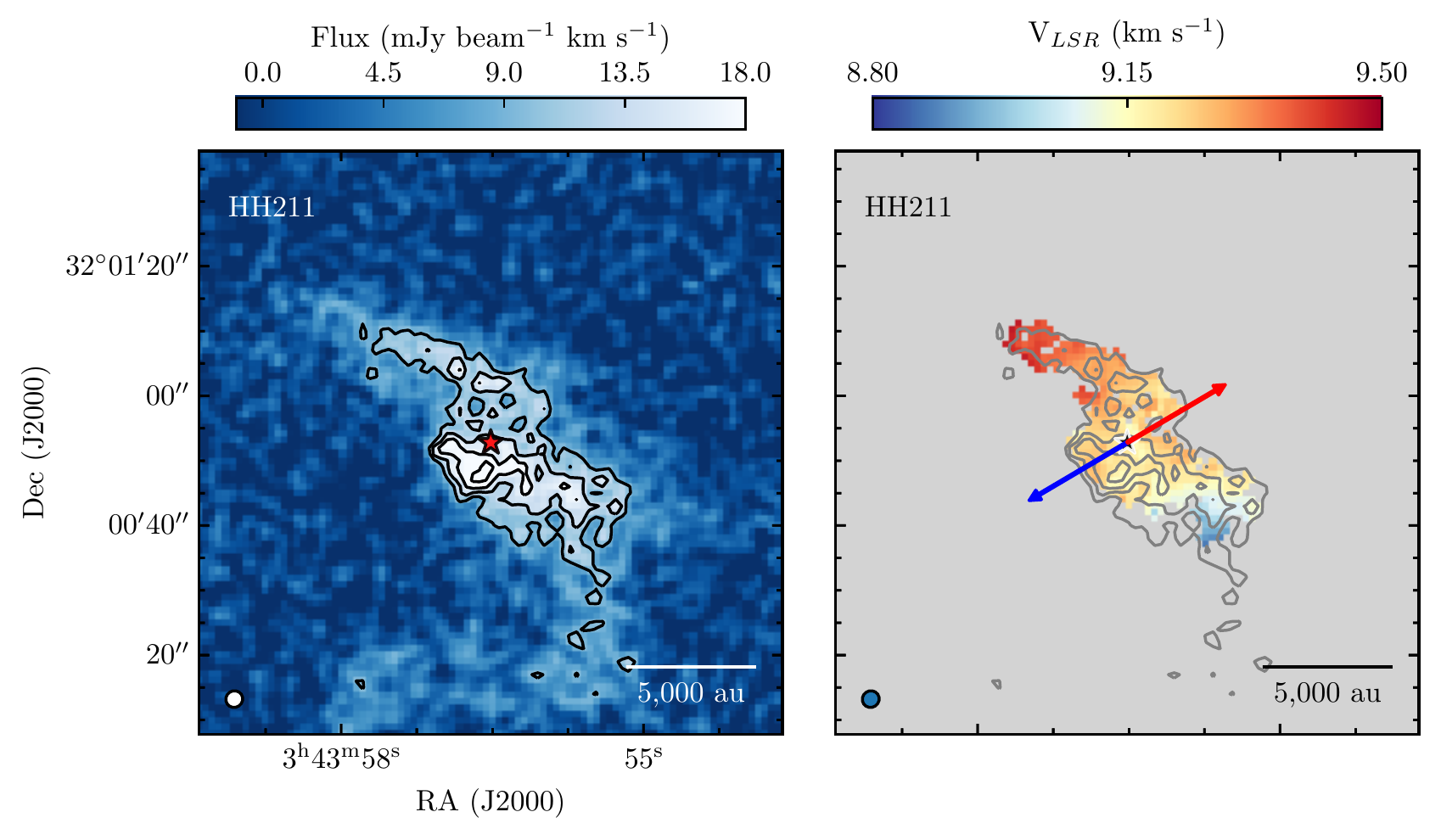}
\caption{ \label{Fig1:HH211}
{Left:} Integrated intensity map of the \ammo(1,1) line. 
{Right:} Velocity field for HH-211 derived from fitting the \ammo(1,1) line.
Outflow orientation is shown by the arrows. 
Contours are shown starting at 5$-\sigma$ with steps of 
2$-\sigma$, where $\sigma=$1.9\,mJy\,beam$^{-1}$\,\kms is the standard deviation measured 
around the emission free region.
Position of the protostar is shown by the star. 
Beam size and scale-bar are shown in the bottom-left and bottom-right corners, respectively.}
\end{center}
\end{figure*}

\begin{figure*}
\begin{center}
\includegraphics[width=\textwidth]{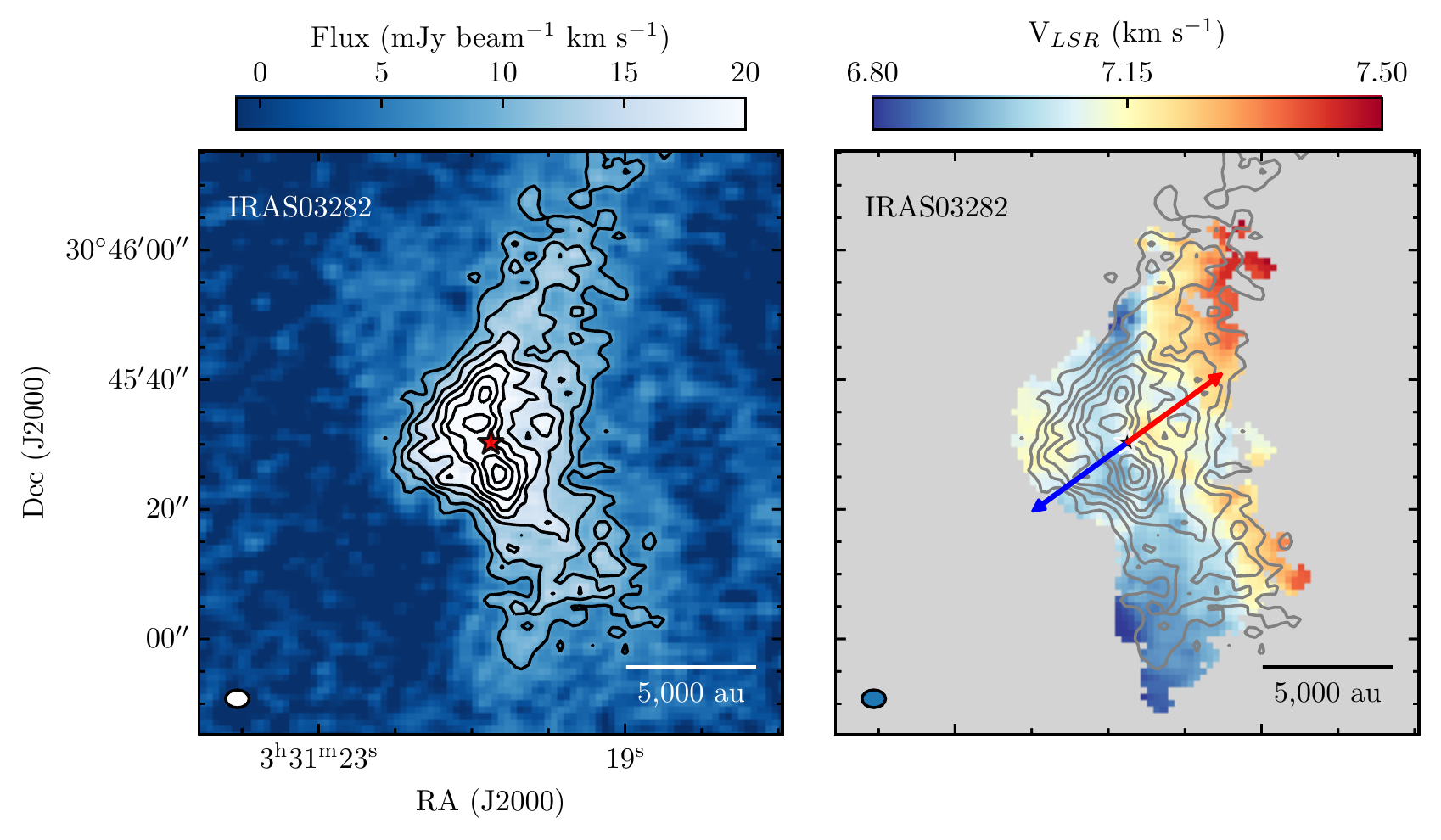}
\caption{ \label{Fig1:IRAS}
Like Fig.\ref{Fig1:HH211}, but for IRAS03282. 
The value of $\sigma$ is 1.6 mJy\,beam$^{-1}$\,\kms.}
\end{center}
\end{figure*}

\begin{figure*}
\begin{center}
\includegraphics[width=\textwidth]{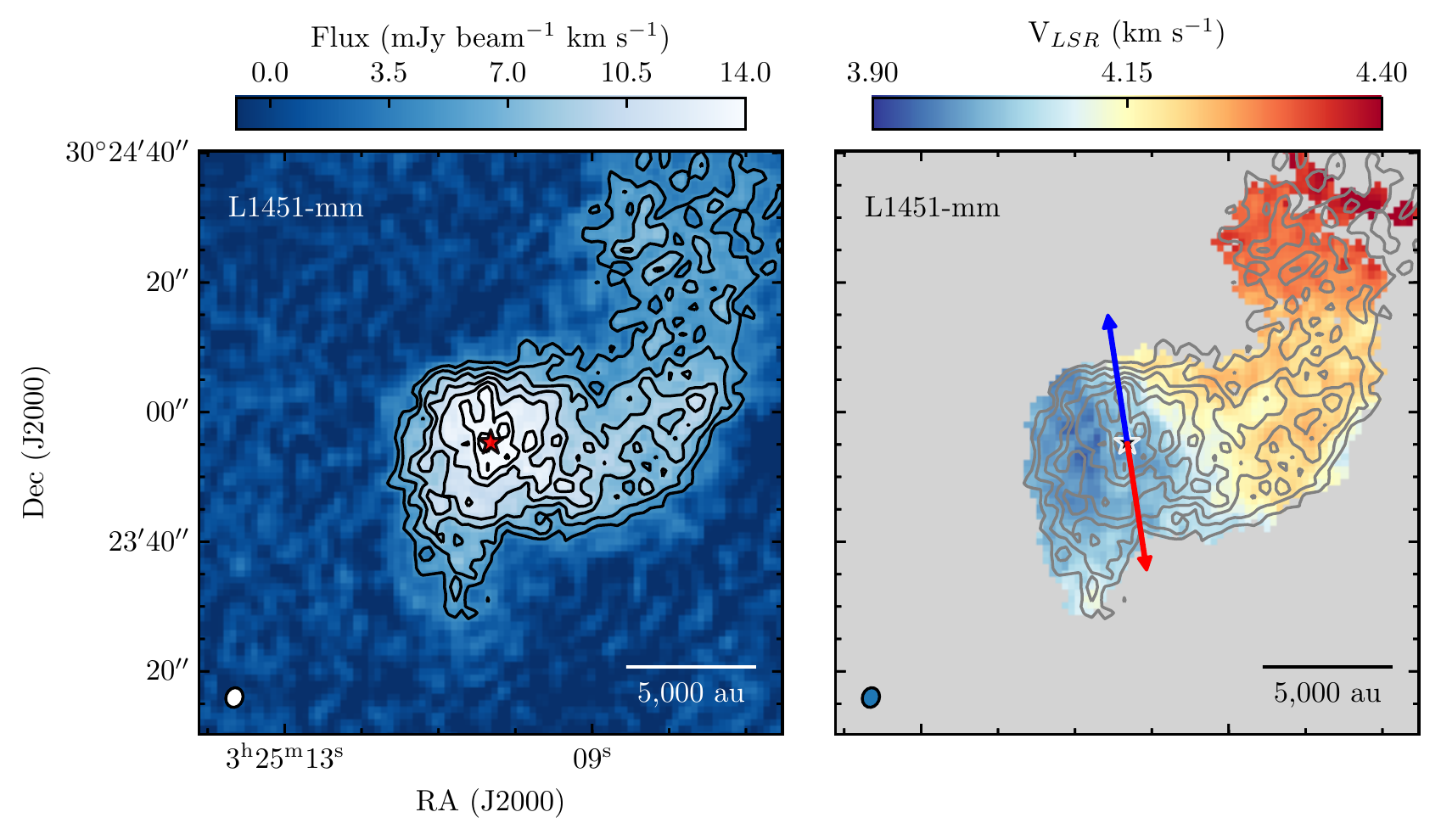}
\caption{\label{Fig1:L1451}
Like Fig.\ref{Fig1:HH211}, but for L1451-mm. 
The value of $\sigma$ is 0.9 mJy\,beam$^{-1}$\,\kms.}
\end{center}
\end{figure*}

For these three sources we fit the \ammo(1,1) line towards all positions with a peak brightness larger than
5$\times$ rms. 
The results are shown in Figure~\ref{Fig1:HH211}, \ref{Fig1:IRAS} and \ref{Fig1:L1451}, where we 
present the integrated intensity and centroid velocity maps. 
The derived velocity dispersion maps are also shown in Figure~\ref{fig-sigmav} (Appendix~\ref{sec:sigmav}), but not used in the analysis.

In the case of source HH211 and L1451-mm the centroid velocity maps show a clear velocity gradient 
in a direction perpendicular to the observed outflow. 
However, in the case of IRAS03282 the centroid velocity map also displays a velocity gradient 
along the outflow direction \citep{Tobin_2011}, which creates a larger spread in the value of the 
specific angular momentum at any given rotation radius.
Nevertheless, since the  effect of the outflow appears to be present in \emph{both} red and blue outflow components, then 
it is possible to derive the average rotational velocity, although with a larger uncertainty.

\subsection{Specific angular momentum radial profile}
The specific angular momentum at a distance $r$, is defined as $j(r)=r \times \delta v$, 
where $\delta v$ is the relative velocity of the gas with respect to the center of mass. 
In the case of symmetry around an axis, then the specific angular momentum can be described as, 
$j(r) = R_{rot} V_{rot}$, 
where $R_{rot}$ is the rotation radius in cylindrical coordinates (also called impact parameter) and 
$V_{rot}$ is the rotational velocity around its axis of symmetry.
In this case, the derived $V_{rot}$ from the observations is similar to the brightness weighted centroid velocity, which corresponds to the mass weighted average velocity in simulations or theoretical calculations.
Therefore, if we determine the rotational velocity at a given rotation radius then we can 
derive the specific angular momentum. 
A similar analysis was performed by \cite{Zhang2018-j} with numerical simulations, showing that the total specific angular momentum could be estimated.
In the case of lines that are not very optically thick, the rotational velocity will correspond to the 
relative centroid velocity of the line along the line-of-sight  ($V_{LSR}$) with 
respect of the core center \citep{Tanner_2010}. 
A similar approach has been used to determine the velocity profile in disks \citep{Murillo_2013,Lindberg2014,Harsono_2014,Harsono_2015}.
In the case of a protostellar core, we use the known YSO position and outflow orientation (see Table~\ref{table:yso}) 
to calculate the rotation radius.
The YSO central velocity is estimated as the velocity of the dense gas, as probed using \ammo, 
at the position of the YSO.
However, since we are not modelling the more complex dynamics involved  closer to the disk scale 
(disk rotation and infall), then 
we discard all determinations of the specific angular momentum at distances smaller 
than the beam major axis.

\begin{figure}[]
\begin{center}
\includegraphics[width=1\columnwidth]{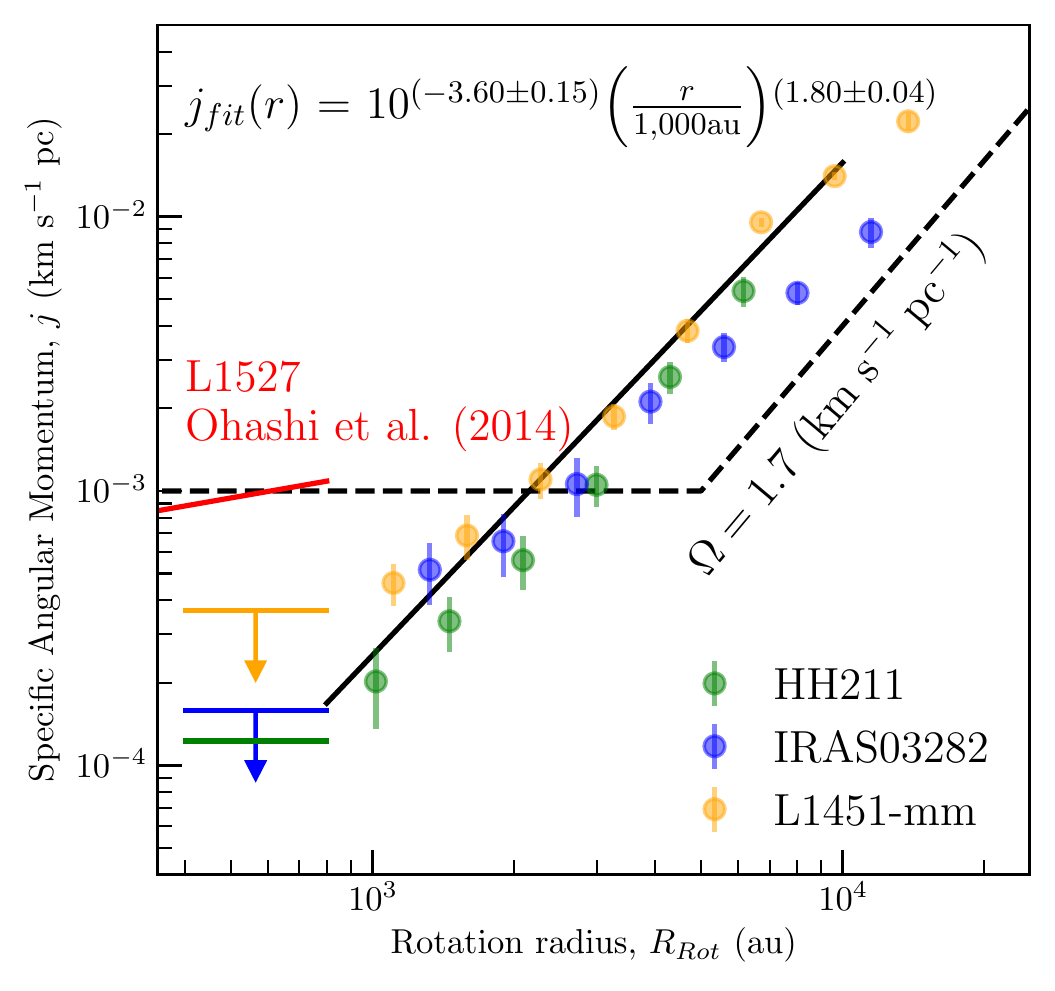}
\caption{\label{Fig:summary}
Radial profile of the specific angular momentum ($j=R_{rot} V_{rot}$) for the three sources studied: 
IRAS03282, HH211, and L1451-mm in blue, green, and yellow, respectively. 
Solid black line shows the best-fit power-law relation to the data between 800 and 10,000\,au. 
The dash curve shows the previously proposed specific angular momentum profile \citep{Ohashi1997,Belloche_2013}. 
The solid red line shows the best-fit power-law relation to the inner envelope of L1527 \citep{Ohashi_2014}, which is a young protostar with a large disk. 
}
\end{center}
\end{figure}

Since the specific angular momentum should be close to constant for a given rotational radius, 
we determine the average value of the specific angular momentum at different rotational radii bins, 
and the associated uncertainty on the mean value. 
The results are shown in Fig.~\ref{Fig:summary}, with the radial profiles for all three sources, 
which appear almost indistinguishable.
The radial profiles display a distribution consistent with a power-law (see Appendix~\ref{sec:v_r} for the velocity comparison  
instead of the specific angular momentum). 
As a comparison, we also show the profile for a core with solid body rotation down to $5,000$\,au, 
and conserved specific angular momentum within that radius using a dash line \citep{Ohashi1997,Belloche_2013}, 
while also comparing to the best-fit to the high-angular resolution ALMA observations on L1527 that studied the 
inner envelope kinematics \citep{Ohashi_2014}. 
The specific angular momentum derived of L1527 is substantially higher than the one measured here, 
which could be related to either a different initial condition or L1527 is a later evolutionary stage object 
(ambiguous due to  the close to edge-on geometry) than those studied here.

The specific angular momentum profiles of the different sources are consistent with a single power-law.
The resulting minimum least-squared power-law fit to all three sources is
\begin{equation}
\label{eq:fit}
j_{fit}(r) = 10^{-3.60\pm 0.15}\left( \frac{r}{\textrm{1,000\,au}}\right)^{1.80\pm0.04} \textrm{km\,s$^{-1}$\,pc}~.
\end{equation}

Notably, we do not find a break in the power-law down to 1,000\,au, 
although there is a hint for a flattening on the radial profiles of IRAS03282 and L1451-mm. 
Higher resolution observations will be crucial to extend the radial profile down to 
smaller radii where the specific angular momentum could be conserved, and directly observe the flattening of the specific angular momentum profile.

\section{Discussion}

\subsection{Interpretation of the power-law}
The results are clear: the specific angular momentum radial profile is consistent with a single power-law down to 1,000\,au. 
However, the power-law exponent is mid-way between the expected exponents 
for solid body rotation ($j\propto r^{2}$) and for turbulence ($j\propto r^{1.5}$) \citep{Burkert_2000}. 

For solid body rotation, the velocity at distance $r$ is $v(r)=\Omega r$, 
where $\Omega$ is the angular speed, 
and therefore the specific angular momentum is $j_{\rm Solid Body}=\Omega r^2$. 
In the case of a pure turbulent field that follows Larson's relation ($\delta v(r) \propto r^{0.5}$),  
the dense cores kinematics (even at scales of $1,000$\,au) might still 
be influenced by the turbulence of the parental molecular cloud/dense core.

This result suggests that dense cores do not present solid body rotation, as usually assumed. 
It might be an intermediate place in between turbulence dominated and solid body rotation.
A similar exponent is also found in simulations of dense cores \citep{Dib_2010}, once cores 
have evolved and are more gravitationally bound.

\subsection{Are these results consistent with those for Dense cores?}
The previous study of the total specific angular momentum, $J/M$, in dense cores by \cite{Goodman_1993} found that 
\begin{equation}
\frac{J}{M} = 10^{-0.7\pm0.2} \left(\frac{R}{\textrm{1\,pc}}\right)^{1.6\pm 0.2}~,
\end{equation}
where $J$ and $M$ are the total angular momentum and mass of the core, and solid body rotation and uniform density are assumed. 
This relation predicts a specific angular momentum at 1,000\,au of $3.95\times 10^{-5}$\,\kmspc, 
which is only 16\% below the directly measured values reported here.

We derived a more general expression for the total specific angular momentum (see Appendix~\ref{sec:deriveEq}), 
\begin{equation}
\label{eq:J_M}
\frac{J}{M}(R) = \frac{(3-k_p)}{2\,(3+k_j-k_p)} j_0 R^{k_j}  \frac{\sqrt{\pi}\, \Gamma(k_j/2 + 1)}{\Gamma((k_j + 3)/2)}~,
\end{equation}
where $\Gamma(x)$ is the Gamma function, 
under the assumption of a density profile $\rho \propto \rho_0 r^{-k_p}$ and a specific angular momentum 
profile $j(r_{rot}) = j_0 r_{rot}^{k_j}$. 
This more general expression allow us to compare our results with those of \cite{Goodman_1993}.

The first thing to note is that \emph{if the specific angular momentum is described as a single power-law out to the dense core radius,} 
then the exponent of the total specific angular momentum as a function of core radius is the same as 
the specific angular momentum profile (eqn.~\ref{eq:J_M}). 
Although the power-law exponents of these relations are slightly different, 
$1.8\pm0.04$ compared to $1.6\pm0.2$, they are 
consistent within the best fit uncertainties. 
We compare the derived total specific angular momentum for the entire core, $J/M(R)$, 
with the intrinsic specific angular momentum at the same scale, $j_{fit}(R)$, using eqn.~\ref{eq:J_M}. 
In the case of $k_j=1.8$ (the best fit value found here), and for typical density profiles ($k_p=0$, $1.5$, or $2$) 
the  
$J/M(R)$ is $0.43\,j(R)$, $0.31\,j(R)$ and $0.25\,j(R)$, respectively. 
Therefore, our results are within a factor of $1.5-3$
from those reported by \cite{Goodman_1993}. 

\subsection{Implications for largest Keplerian disk radius}
Although we do not find the  region with conserved specific angular momentum, 
if it is conserved from the 1,000\,au scale down 
to the disk formation scales, then we can estimate the maximum possible Keplerian disk size 
using the relation 
\begin{equation}
R_{disk} = \frac{j_e^2}{G\,M_*}~,
\end{equation}
see \cite{Yen_2015}, where $j_e$ is the specific angular momentum of the material that will form the disk,
and $M_*$ is the stellar mass. 
This is a conservative upper limit to the Keplerian disk size, since we are not estimating the total enclosed mass only the stellar mass, and any other mechanism of angular momentum removal would reduce the infant disk radius.

We use our best fit to the angular momentum radial profile at 1,000\,au, eq.(\ref{eq:fit}), 
and estimate an upper limit to $j_e$ of $10^{-3.60}$\,\kmspc.
In the case of  a stellar mass 
of 0.05\,M$_\odot$, which is appropriate for HH211 \citep{Lee_2009,Froebrich_2003} and 
L1451-mm \citep{Pineda_2011,Maureira2017-L1451mm}, 
we estimate an upper limit to the Keplerian disk radius of 60\,au. 
This is consistent with disk radius estimates (including upper limits) for these sources using 
high-angular resolution dust continuum observations (see Table~\ref{table:yso}).

\subsection{Estimating protostellar age}
Some theoretical work relates the value of the constant specific angular momentum 
with the protostellar age \citep{Takahashi2016-j}. 
Again, we assume the specific angular momentum value at 1,000\,au from the best fit, $j_{fit}(r)$ to estimate the 
collapse age of the core for these protostars using equation~26 from \citep{Takahashi2016-j} of 
11\,kyr. 
This is consistent to the accretion age estimated for HH211 and L1451-mm of 
$(1-2)\times10^4$\,yr \citep{Lee_2009,Maureira2017-L1451mm}. 
If this interpretation is correct, then it predicts that more evolved sources (e.g., Class~I) would show a similar exponent 
but a variety of flattening radii depending on the source age, similar to the results from \cite{Yen2017} who assumed
solid body rotation for the envelope.

\subsection{Connecting with numerical simulations}
Both the exponent and normalisation of the specific angular momentum profile derived from our observations 
are substantially different than those 
typically used as initial conditions for disk formation simulations \citep{Ohashi1997,Belloche_2013} or as 
a point of comparison for core properties in global simulations \citep{Jappsen2004-j,Ntormousi2019-Cores,Kuznetsova2019-j}. 
Here, we explore the implications for numerical simulations. 

\begin{figure}
\includegraphics[width=\columnwidth]{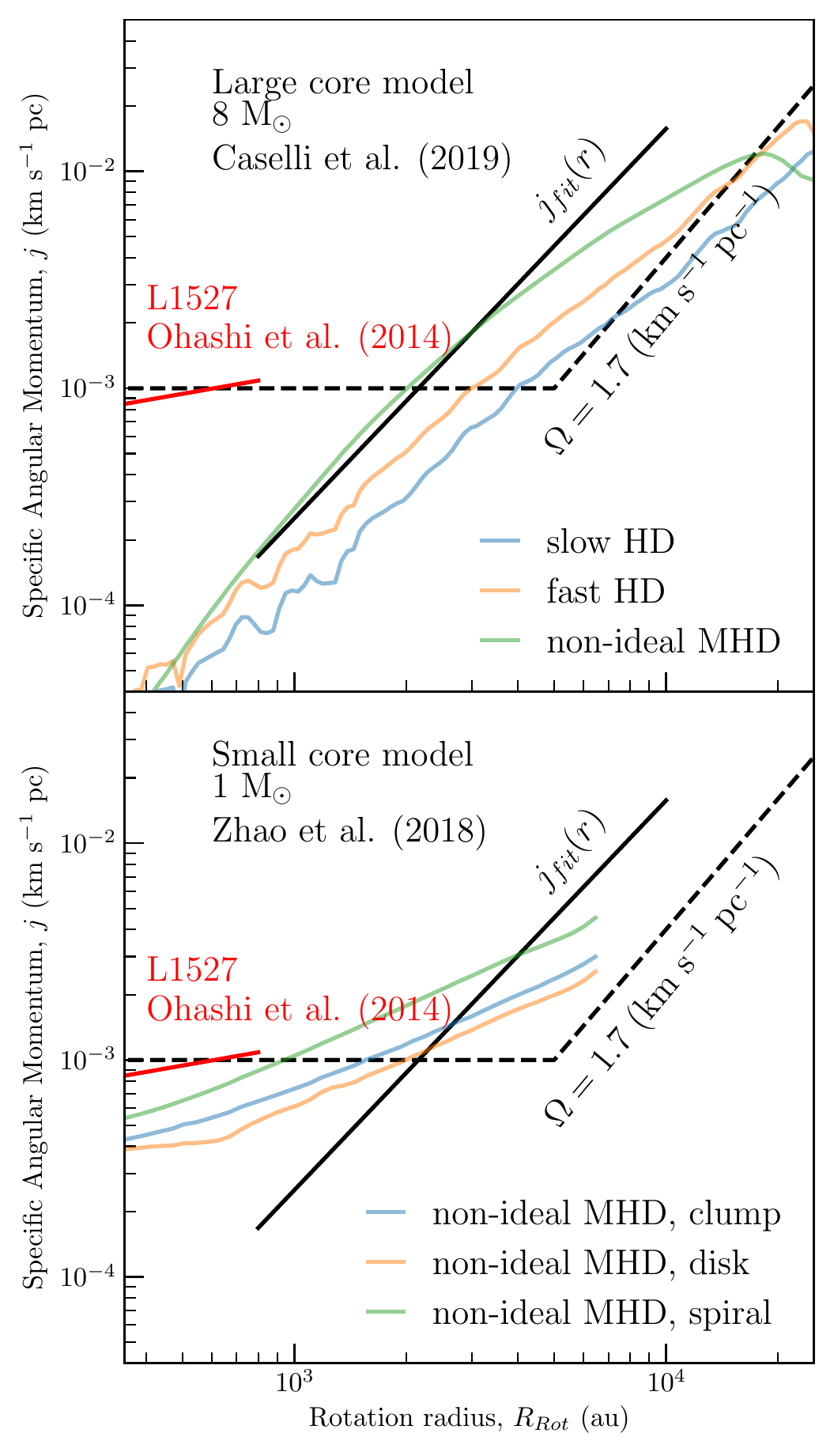}
\caption{Specific angular momentum profile for the different sets of simulations in the large and small core models, top and bottom panels respectively.
Our best fit to the data, $j_{fit}(r)$, is shown with a solid black line; the previously derived specific angular profile \citep{Ohashi1997,Belloche_2013} is shown with a dash black line; the power-law fit to the L1527 inner envelope kinematics \citep{Ohashi_2014} is shown with a red line. 
The non-ideal MHD simulation of the large prestellar core provides the best match to the observations of the Class~0 sources, 
while the non-ideal MHD simulations of a small protostellar core are closer to the observed kinematics of the (probably more evolved) L1527 source.
\label{fig-simu}} 
\end{figure}

We use two sets of simulations already published in previous work \citep{Zhao2018,Caselli2019-L1544} to explore the global trends of the specific angular momentum radial profile, 
but these simulations are not expected to be a perfect match to the presented data.
The two simulations sets are split as large and small core models, and for each simulation we calculate the average specific angular momentum profiles for a given snapshot.
The general descriptions of the simulations are the following: 
\paragraph{(a)}
The large core models correspond to a 8.1\,\msun dense core with a radius 50,000\,au numerical simulations using HD and non-ideal magneto-hydro-dynamical (MHD) simulations showing the snapshot where the central density is 2$\times10^6$\,cm$^{-3}$, which are aimed at reproducing the properties of the L1544 prestellar core \citep[more details in][]{Caselli2019-L1544}.  
Two HD simulations are run: ``slow HD'' and ``fast HD'' with angular speeds of 2.5$\times10^{-14}$\,s$^{-1}$ and 
4$\times10^{-14}$\,s$^{-1}$, respectively, with the``fast HD'' simulation rotating close to the maximum speed that would allow collapse ($\Omega_{max}$=6$\times10^{-14}$\,s$^{-1}$ in the case of $\beta$=0.5 and solid body rotation). 
The MHD simulation is run with angular speed of 4$\times10^{-14}$\,s$^{-1}$, magnetic field of $B_0$=14.8\,$\mu$G, and a mass-to-flux ratio $\lambda$ of $\sim$1.

\paragraph{(b)}
The small core models correspond to a 1\,\msun dense core with a 6,666\,au radius, using non-ideal MHD simulations with different angular speed rotation and magnetic field strength as initial conditions \citep[more details in][]{Zhao2018}.
Here we consider the snapshot where the star and disk total mass is $\sim$0.1\,\msun, 
aimed at reproducing protostellar systems in the early stages of evolution (close to Class~0). 
The ``disk'' and ``spiral'' simulations have a magnetic field of 42.5\,$\mu$G ($\lambda\sim2.4$), while 
the ``clump'' simulation has a magnetic field of 21.3\,$\mu$G  ($\lambda\sim4.8$). 
The ``disk'' and ``clump'' simulations have an angular speed of $10^{-13}$\,s$^{-1}$, while 
the ``spiral'' simulation has an  angular speed of $2\times10^{-13}$\,s$^{-1}$.

Figure~\ref{fig-simu} compares the best fit to the data and the profiles from the simulations. 
Top panel of Figure~\ref{fig-simu} shows the profiles for the large core model that has not formed a protostar yet, in which the non-ideal MHD simulation presents a smoother profile and with a slope and amplitude in the inner 4,000\,au which is much closer to the best fit to observations than the HD simulations. The main reason for this increased specific angular momentum in the non-ideal MHD simulation compared to the HD ones, is that the collapse takes longer in the non-ideal MHD case to reach the same central density. This keeps  the angular momentum in the outer parts fo the dense core for a longer period of time, which generates an increased specific angular momentum at intermediate scales (700 -- 4,000 au).
The bottom panel of Figure~\ref{fig-simu} shows the profiles for the small core model that already formed a disk, which are much flatter than our best fit from observations. 
However, all these models are consistent with the result for L1527 shown with a red line \citep{Ohashi_2014}, since each simulation will evolve with time in such a way that the radial profile does not change the slope, but it does change the normalisation.
These results suggests that the Class~0 sources here observed are indeed very young, and their dense core kinematics might still resemble that of the parental dense core. 
On the other hand, the kinematics of the (possibly more evolved) L1527 is better reproduced by simulations which already formed a protostar and disk system. 
This highlights the need to properly determine the probable evolution of the specific angular momentum as a function of source evolutionary stage.

\section{Summary and Conclusions}

We analysed VLA \ammo(1,1) data for 3 of the youngest protostellar sources in the Perseus molecular cloud to determine the dense gas kinematics of the material involved in the disk formation process. 
Our main results are summarised below.

\begin{itemize}
\item We determine the specific angular momentum radial profile for three YSOs and 
find that is consistent among these three sources:
\[ j(r) = 10^{-3.60\pm0.15}\left( \frac{r}{1,000 \textrm{\,au}}\right)^{1.80\pm 0.04}~\kmspc~,\]
and find that the region of conserved specific angular momentum must be smaller than 1,000\,au radius. 
\item We determine expressions for the specific angular momentum of the entire core under 
a more general condition, $j(r) \propto r_{rot}^{k_j}$, than solid body rotation.
\item The specific angular momentum radial profile normalisation is consistent with previous determinations 
of the total specific angular momentum based on dense core solid body rotation fits, 
but only after we correct for the non-solid body rotation nature of the cores.
\item If the specific angular momentum is conserved inwards of 1,000\,au, then the 
largest disk towards these low-mass YSOs ($M_*\sim$0.05\,\msun) is 60\,au in radius, which is consistent 
with previous high-angular resolution studies focused on the disk properties.
\item We compare to already published numerical simulations for a large prestellar dense core and a small protostellar dense core. 
The comparison shows that the non-ideal MHD simulation of a prestellar core provides a  good match to the observed profile, and better than HD simulations. 
The protostellar simulations show profiles more similar to that of L1527. 
This suggests a magnetic field regulated dynamical evolution between pre-stellar and protostellar cores.
\end{itemize}

These results show the potential for constraining the disk formation process by connecting to the knowledge of the 
parental dense core properties.  
Future observations might reveal possible variations with environment and evolution on the specific angular momentum profile.

\acknowledgments
We thank the anonymous referee for constructive comments that helped to improve this manuscript. J.E.P thanks Alyssa A. Goodman, Hope H.-H. Chen, Anaelle J. Maury, Stella Offner, Ralf Klessen, Kengo Tomida, and Nagayoshi Ohashi for valuable discussions.
The Expanded Very Large Array is operated by the National Radio
Astronomy Observatory. The National Radio Astronomy Observatory
is a facility of the National Science Foundation, operated
under cooperative agreement by Associated Universities, Inc.
J.E.P. and P.C. acknowledge the financial support of the European Research Council (ERC; project PALs 320620). 

\facility{EVLA}
\software{CASA \citep{2007ASPC..376..127M}, Matplotlib \citep{Hunter:2007}, APLpy \citep{APLpy2012}, PySpecKit \citep{2011ascl.soft09001G}}

\appendix

\section{Velocity dispersion maps\label{sec:sigmav}}
\begin{figure}[h]
\includegraphics[width=0.333\textwidth]{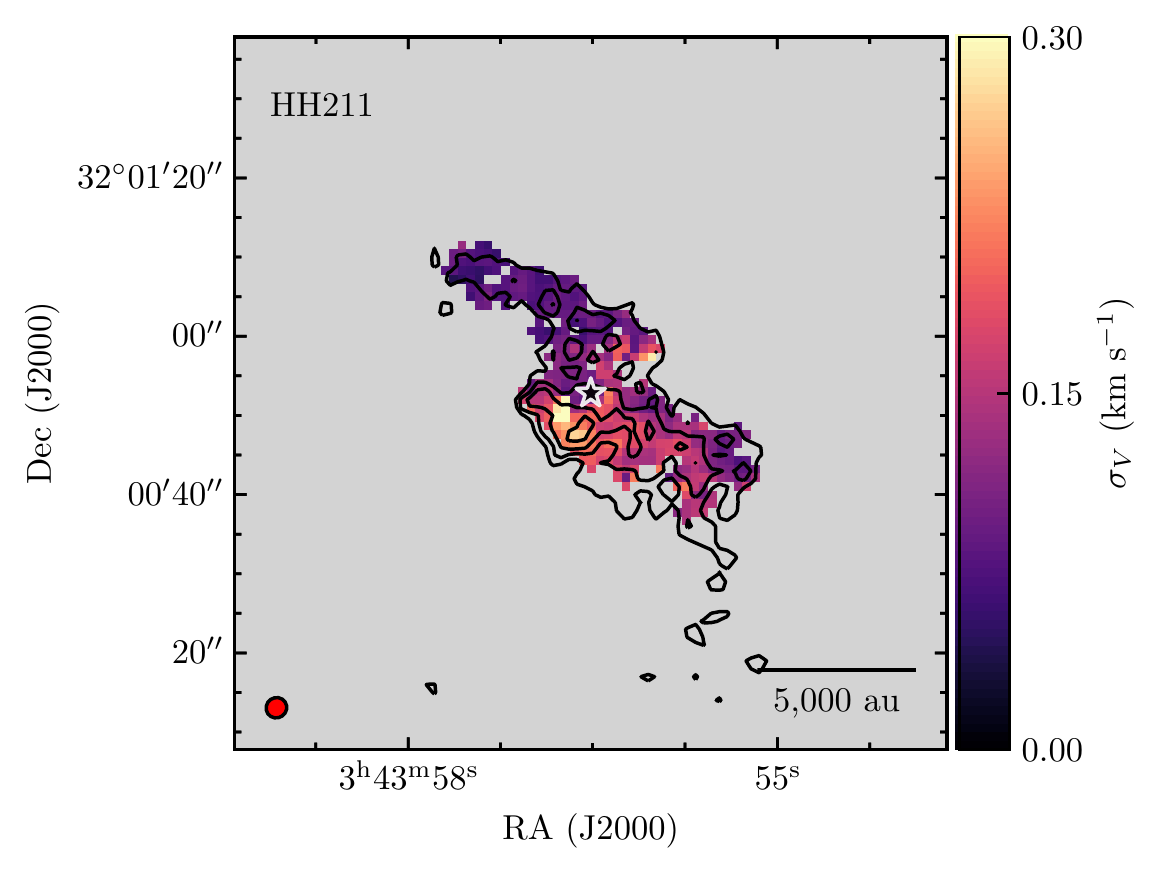}
\includegraphics[width=0.333\textwidth]{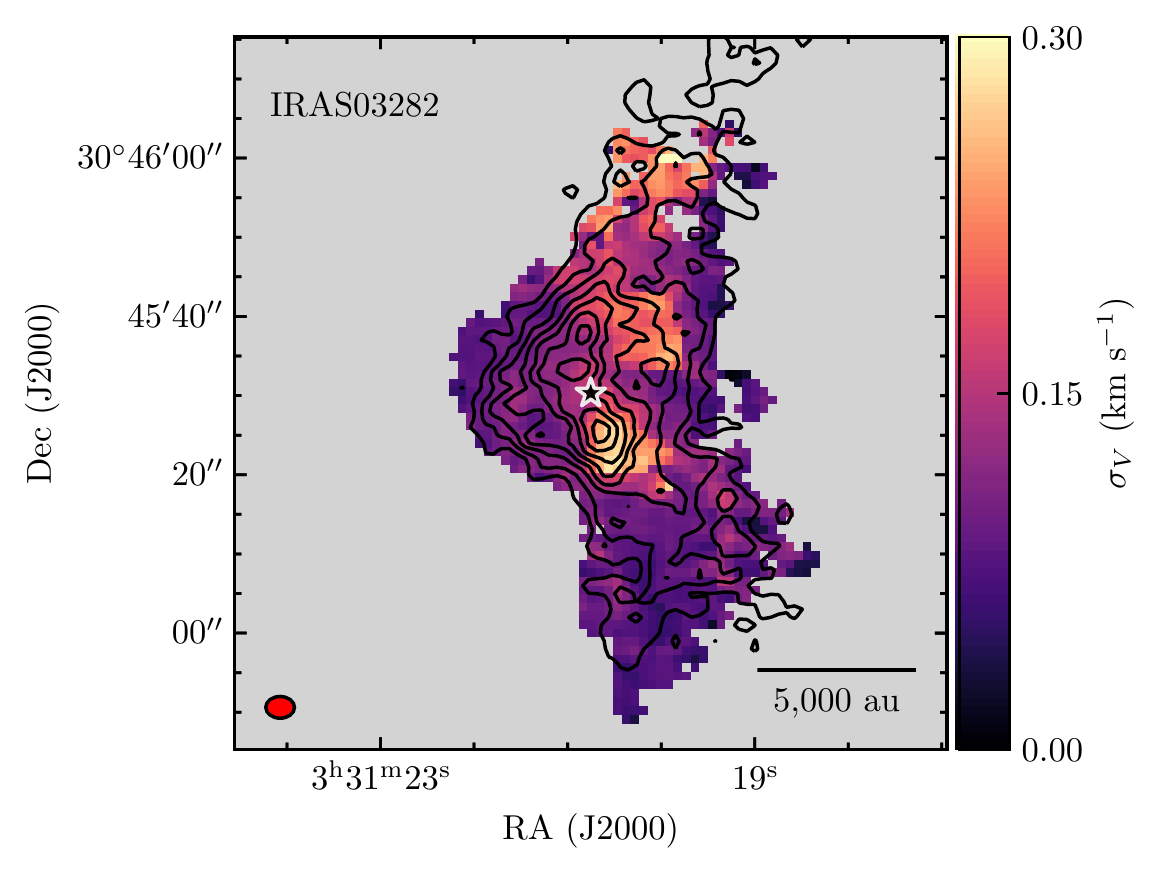}
\includegraphics[width=0.333\textwidth]{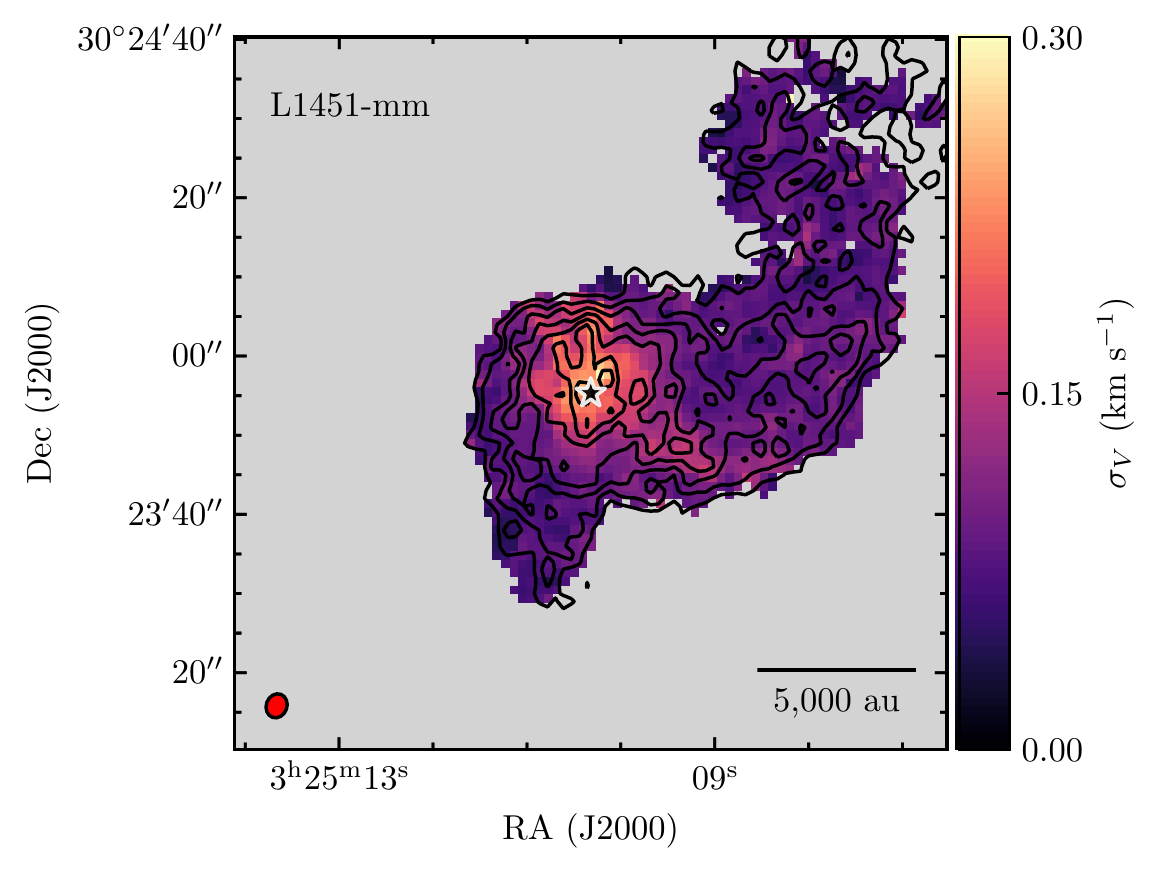}
\caption{Velocity dispersion maps for all three sources: HH-211, IRAS03282, and L1451-mm in the left, middle, and  right panels, respectively. 
Beam size and scale bar are shown in bottom left and right corners, respectively.
\label{fig-sigmav}}
\end{figure}

The velocity dispersion maps (without removal of channel width response) for all three sources are shown in Figure~\ref{fig-sigmav}. 
As reported in previously, these velocity dispersions correspond to non-thermal velocity dispersions below the sonic sound speed (Mach number $<0.5$), but with an increase towards to the protostar position, which is most likely due to protostellar feedback.

\section{Rotational velocity as a function of radius\label{sec:v_r}}
\begin{figure}[h]
\centering
\includegraphics[width=0.5\textwidth]{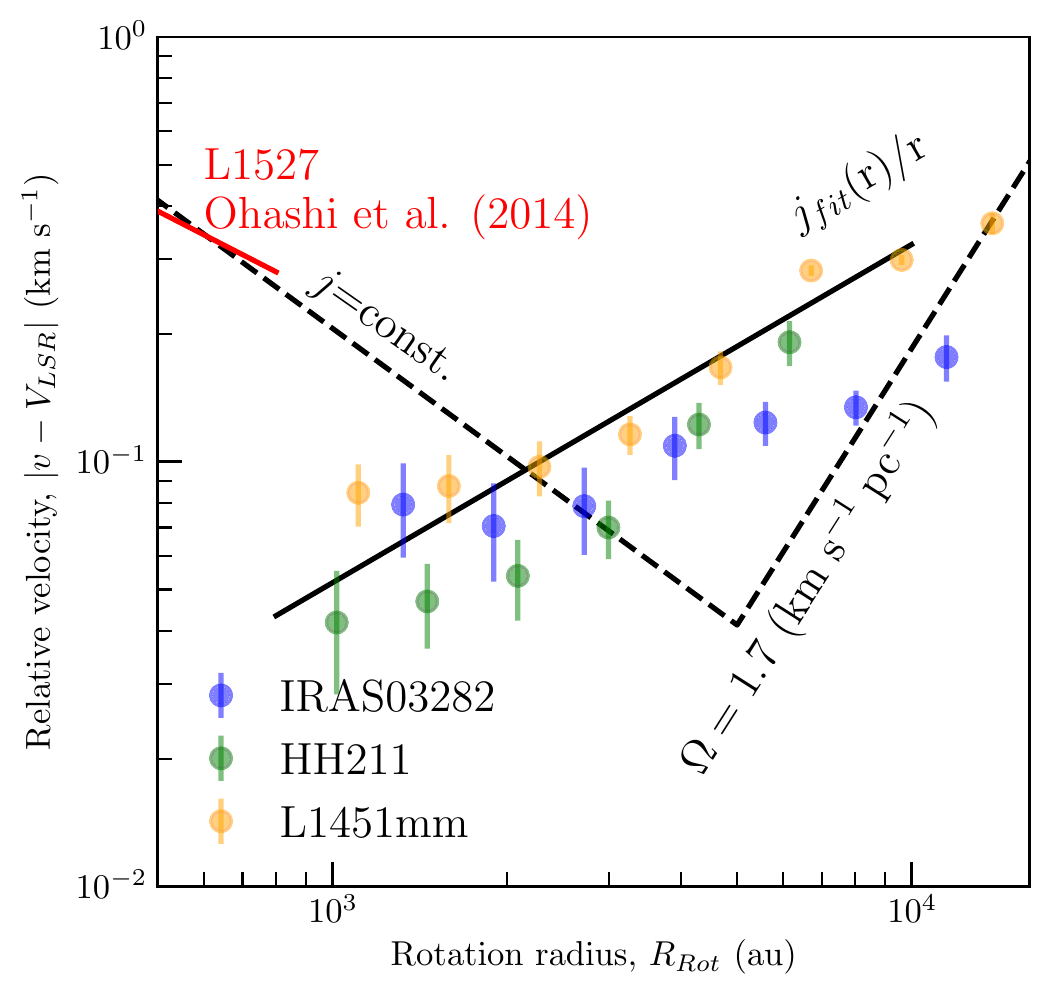}
\caption{\label{fig:v-r}
Rotational velocity as a function of radius for all three sources observed, measured using the relative velocity measurement. 
Similar to Fig.~\ref{Fig:summary}, the black solid line is the expected rotational velocity derived from the specific angular momentum best fit, $V_{fit}(r) = j_{fit}(r)/r$,
while the proposed rotational velocity profile by \cite{Ohashi1997} and \cite{Belloche_2013} is shown by the dashed line. 
The data show a clear correlation with radius which is well described by $V_{fit}(r)$, 
while the dashed line profile poorly describes the data. 
The best fit rotational velocity profile derived for L1527~IRS by \cite{Ohashi_2014} is shown by the red line.}
\end{figure}

A reasonable concern is that the calculation of $j(r)=R_{rot}\,V_{rot}$ could be mostly driven by $R_{rot}$, instead of  a real radial dependence. 
A direct check for this possible issue is done by comparing the average velocity as a function of radius, using the same radial bins as the average specific angular momentum as a function of radius. 
Notice, that for this calculation we use the actual relative velocity derived from the data.
We compare the results for all three sources in Fig.~\ref{fig:v-r} against the predicted rotational velocity as a function of radius for the specific angular momentum best fit and for the proposed 
profile by \cite{Ohashi1997} and \cite{Belloche_2013}. 
The data are better described by a single power-law, without evidence for a transition to a 
conserved angular momentum regime, as previously proposed. 
This comparison confirms our results from fitting the specific angular momentum.

\section{Properties as function of radius\label{sec:deriveEq}}
Here we show how to determine the total angular momentum ($J$) and specific angular momentum ($J/M$) in the case of differential rotation and/or a radial density profile.

Assuming that the density profile is spherically symmetric and that it can be described 
by a single power-law, depending only on the distance to the central protostar:
\begin{equation}
\rho = \rho_0 r^{-k_p}~.
\end{equation}
Naturally, the total enclosed mass can be described as 
\begin{equation}
M(<R) = \frac{4\pi}{(3-k_p)} \rho_0 R^{3-k_p}~,
\end{equation}
for $k_p\ne 3$.

Also, assuming that the specific angular momentum can be described in spherical coordinates as
\begin{equation}
j(r,\theta) = j_0 ( r \sin \theta)^{k_j}~,
\end{equation}
where $k_j=2$ represents solid body rotation, 
the total angular momentum within a radius $R$ is
\begin{eqnarray}
J(R) &=& \int_{Volume} j(r,\theta) \rho(r) d\Omega r^2 dr \\
&=& \int_0^{2\pi}\int_{0}^{\pi}\int_0^R j_0 ( r \sin \theta)^{k_j}\, \rho_0 r^{-k_p}\, \sin\theta r^2\, dr d\theta d\phi\\
&=& 2\pi \int_{0}^{\pi} \sin^{k_j+1} \theta d\theta \int_0^R \rho_0 j_0 r^{2+k_j-k_p} \,dr \nonumber \\
&=& \frac{2\pi}{(3+k_j-k_p)}\rho_0 j_0 R^{(3+k_j-k_p)} \int_{0}^{\pi} \sin^{k_j+1} \theta d\theta~,
\end{eqnarray}
for $(k_p-k_j)\ne 3$.

The total specific angular momentum can be written as
\begin{equation}
\frac{J}{M}(R) = \frac{(3-k_p)}{2(3+k_j-k_p)} j_0 R^{k_j} 
\int_{0}^{\pi} \sin^{k_j+1} \theta d\theta~.
\end{equation}
The defined integral can be written as 
\begin{eqnarray}
\int_{0}^{\pi} \sin^{\alpha+1} x dx &=&  \frac{\sqrt{\pi}\, \Gamma(\alpha/2 + 1)}{\Gamma((\alpha + 3)/2)}
\end{eqnarray} 
for $\alpha>-2$, and 
where $\Gamma(x)$ is the Gamma function.
Therefore, 
\begin{equation}
\frac{J}{M}(R) = \frac{(3-k_p)}{2\,(3+k_j-k_p)} j_0 R^{k_j}  \frac{\sqrt{\pi}\, \Gamma(k_j/2 + 1)}{\Gamma((k_j + 3)/2)}~,
\end{equation}
equivalent to equation~\ref{eq:J_M}.

The rotational energy is calculated as,
\begin{eqnarray}
E_{rot}(R) &=& \int \frac{1}{2} v_{rot} (r,\theta)^2 \rho(r) dV \\
&=& \pi \int \left(j_0 (r \sin \theta)^{k_j-1}\right)^2 \rho_0 r^{-k_p} \sin\theta r^2 dr d\theta \nonumber\\
&=& \pi  \int j_0^2 (r \sin \theta)^{2\,k_j-2} \rho_0 r^{2-k_p} \sin\theta dr d\theta  \nonumber\\
&=& \pi \rho_0  j_0^2  \int_0^R  r^{2k_j-k_p}  dr  \int_0^\pi  \sin^{2\,k_j-1} \theta  d\theta  \nonumber\\
&=& 
\frac{\pi \rho_0  j_0^2 }{(2k_j-k_p+1)} R^{2k_j-k_p+1}
\int_0^\pi  \sin^{2k_j-1} \theta  d\theta   \nonumber\\
&=& 
\frac{\pi \rho_0  j_0^2 }{(2k_j-k_p+1)} R^{2k_j-k_p+1}
\frac{\sqrt{\pi}\, \Gamma(k_j)}{\Gamma(k_j +1/2)}
 \nonumber\\
 &=&  
\frac{(3-k_p)}{(2k_j-k_p+1)}  \frac{M R^{2(k_j-1)} j_0^2 }{4}
\frac{\sqrt{\pi}\, \Gamma(k_j)}{\Gamma(k_j +1/2)}~.
\end{eqnarray}

The total gravitational energy is 
\begin{eqnarray}
U &=& -\frac{(4\pi\rho_0)^2 G}{(3-k_p)(5-2k_p)} R^{5-2k_p}\\
&=& -\frac{(3-k_p)}{(5-2k_p)}\frac{G M^2}{R}~.
\end{eqnarray}

The ratio of the rotational and gravitation energy, $\beta=E_{rot}/|U|$, is
\begin{eqnarray}
\beta 
&=& 
\frac{j_0^2 }{(2k_j-k_p+1)} R^{2k_j+ k_p -4}
\frac{(3-k_p)(5-2k_p)}{16\pi\rho_0 G} 
\frac{\sqrt{\pi}\, \Gamma(k_j)}{\Gamma(k_j +1/2)}
 \\
&=& 
\frac{(3-k_p)}{(2k_j-k_p+1)}  \frac{M R^{2(k_j-1)} j_0^2 }{4}
\frac{\sqrt{\pi}\, \Gamma(k_j)}{\Gamma(k_j +1/2)}
\frac{(4\pi\rho_0)^2 G}{(3-k_p)(5-2k_p)} R^{5-2k_p} \\
&=& 
\frac{(5-2k_p)}{(2k_j-k_p+1)}  
\frac{R^{2k_j-1} j_0^2 }{4 G M}
\frac{\sqrt{\pi}\, \Gamma(k_j)}{\Gamma(k_j +1/2)}~.
\end{eqnarray}

\bibliography{bibliography}

\end{document}